\documentclass[aps, pra, reprint, superscriptaddress, nofootinbib]{revtex4-2}

\usepackage{amsmath}
\usepackage{amssymb}
\usepackage{graphicx}
\usepackage{hyperref}
\usepackage{color}
\usepackage{siunitx}
\usepackage{xcolor}
\usepackage{soul}
\usepackage{bm}

\allowdisplaybreaks

\begin{document}

\title{Prospects for measuring the electron's electric dipole moment with polyatomic molecules in an optical lattice}

\author{Roman Bause}
\thanks{Present address: Menlo Systems GmbH, 82152 Martinsried, Germany.}
\affiliation{Van Swinderen Institute for Particle Physics and Gravity, University of Groningen, The Netherlands}
\affiliation{Nikhef, National Institute for Subatomic Physics, Amsterdam, The Netherlands}

\author{Nithesh Balasubramanian}
\author{Ties Fikkers}
\author{Eifion H. Prinsen}
\affiliation{Van Swinderen Institute for Particle Physics and Gravity, University of Groningen, The Netherlands}
\affiliation{Nikhef, National Institute for Subatomic Physics, Amsterdam, The Netherlands}

\author{Kees Steinebach}
\affiliation{LaserLaB, Vrije Universiteit Amsterdam, The Netherlands}

\author{Arian Jadbabaie}
\affiliation{Department of Physics, Massachusetts Institute of Technology, Cambridge, Massachusetts 02139, United States of America}
\author{Nicholas R. Hutzler}
\affiliation{Division of Physics, Mathematics, and Astronomy, California Institute of Technology, Pasadena, CA 91125, United States of America}

\author{I. Agust\'{\i}n Aucar}
\affiliation{Van Swinderen Institute for Particle Physics and Gravity, University of Groningen, The Netherlands}
\affiliation{Nikhef, National Institute for Subatomic Physics, Amsterdam, The Netherlands}
\affiliation{Instituto de Modelado e Innovaci\'on Tecnol\'ogica (UNNE-CONICET), Facultad de Ciencias Exactas y Naturales y Agrimensura, Universidad Nacional del Nordeste, Corrientes, Argentina}

\author{Luk\'{a}\v{s} F. Pa\v{s}teka}
\affiliation{Van Swinderen Institute for Particle Physics and Gravity, University of Groningen, The Netherlands}
\affiliation{Nikhef, National Institute for Subatomic Physics, Amsterdam, The Netherlands}
\affiliation{Department of Physical and Theoretical Chemistry, Comenius University,  Bratislava, Slovakia}

\author{Anastasia Borschevsky}
\author{Steven Hoekstra}
\email{email: s.hoekstra@rug.nl}
\affiliation{Van Swinderen Institute for Particle Physics and Gravity, University of Groningen, The Netherlands}
\affiliation{Nikhef, National Institute for Subatomic Physics, Amsterdam, The Netherlands}

\date{\today}

\begin{abstract}	
We present the conceptual design of an experiment to measure the electron's electric dipole moment (eEDM) using $^{138}$BaOH molecules in an optical lattice. The BaOH molecule is laser-coolable and highly sensitive to the eEDM, making it an attractive candidate for such a precision measurement, and capturing it in an optical lattice offers potentially very long coherence times. We study possibilities and limitations of this approach, identify the most crucial limiting factors and ways to overcome them. The proposed apparatus can reach a statistical error of $10^{-30}\,e\,$cm by measuring spin precession on a total number of $5 \times 10^9$ molecules over a span of 120 days.
\end{abstract}

\maketitle

\section{Introduction}
Over the last two decades, the experimental determination of the electric dipole moment of the electron (eEDM) has been improved by three orders of magnitude, pushing the upper bound down to $4.1\times 10^{-30}\,e\,\text{cm}$~\cite{Regan_2002, Hudson_2011, Eckel_2013, ACME_2013, Cairncross_2017, ACME_2018, Roussy_2023}. This enormous sensitivity is enabled by leveraging cold polar molecules containing a heavy element and large electron spin density at the nucleus. Their well-defined transitions allow experimentalists to measure small frequency shifts caused by the dipole moment aligning in a field, which can be done with extreme precision. This category of molecules offers energy levels which are very sensitive to the eEDM, with predicted enhancement factors on the order of $W_\mathrm{d} \sim 10^{24}\,h\,$Hz/($e\,$cm)~\cite{Haase_2021, Chamorro_2022}. 

The eEDM is considered interesting because of its connection to the violation of $CP$ symmetry~\cite{Safronova_2018, Chupp_2019}. There are multiple reasons why it is believed that the Standard Model of particle physics does not fully describe $CP$ violation in nature, such as the strong $CP$ problem~\cite{Kim_2010} and the matter-antimatter asymmetry in the universe~\cite{Dine_2003}. In the Standard Model, the EDM of a bare electron is predicted to be extremely small, on the order of $10^{-40}\,e\,\text{cm}$, because it arises only at the fourth order of perturbation theory~\cite{Yamaguchi_2020}. What is measured in experiments is the EDM of a molecule, where there exist additional $CP$-odd  Standard Model electron-nucleon interactions that manifest an effective EDM at the level of $10^{-35}\,e\,\text{cm}$~\cite{Ema_2022}. These sources of $CP$ violation can be disentangled by combining measurements on different atoms or molecules and electronic-structure calculations. 

If there were additional sources of $CP$ violation in nature, these would in many cases couple to the electron at one-loop level, leading to a much larger effective EDM~\cite{Chupp_2019}. Indeed, the present experimental upper bound for the free-electron EDM already rules out some beyond-Standard-Model theories at energy scales beyond the range of the largest colliders~\cite{DeMille_2017, Cesarotti_2019, Roussy_2021}. 

Here we propose a new apparatus which can reach state-of-the-art sensitivity and opens up avenues for further improvements. The two main ingredients to our proposal are: (1) the use of barium monohydroxide ($^{138}$BaOH), which is laser-coolable and allows suppression of certain systematic effects using internal comagnetometry, and (2) holding the molecules in an optical lattice, enabling interrogation times on the order of a second. Building on previous proposals \cite{Hutzler_2020, Fitch_2020, Cairncross_2019, Augenbraun_2021a, Anderegg_2023}, we study experiment designs in more detail. With better understanding of technical limitations and ways to overcome them, we make a step towards constructing such an experiment. We propose the use of a power build-up cavity to construct a deep optical lattice with high polarization purity, using mixed samples of both projection states in each run of the experiment for co-magnetometry, and reducing the effective magnetic moment of these eEDM states by tuning an electric field.

The paper is organized as follows. In Section~\ref{sec-req} we first introduce the general concept of EDM experiments, followed by an overview of the effects which limit the achievable precision. The main purpose of this section is to develop understanding of the advantages and challenges which come with performing an EDM measurement in an optical dipole trap. This includes both the molecule-light interaction itself and other effects like molecule lifetime in the trap. In Section \ref{sec:calculations} we present theoretical predictions of certain quantities which are needed to plan the experiment. The technical details of the planned experiment are presented in Section~\ref{sec-implementation}. Here, we first discuss how to obtain a suitable molecule sample, followed by a study of the dipole trap used for the EDM measurement. We further present how to create the fields with the required stability, and conclude this section with details of the envisaged spin precession readout and molecule detection. We complete the paper with Conclusion (\ref{sec-conclusion}) and Outlook (\ref{sec-outlook}) sections.

\section{Requirements for the planned experiment}\label{sec-req}

\subsection{Outline of the concept}\label{sec-outline}
The basic concept of measuring the molecule EDM is the same for all recent experiments: a sample of well-suited molecules is created and brought into a superposition of two states with equal but opposite magnetic quantum numbers, for example $|m_F = +1\rangle + |m_F= -1\rangle$ of the $F=1$ hyperfine level. These molecules then undergo spin precession in a volume with well-controlled $E$ and $B$ fields, which are either parallel or antiparallel to each other. If there is a finite EDM, then the precession frequencies will be different in these two cases. If no difference is visible within the measurement precision, an upper bound on the eEDM can be extracted. To avoid systematic effects from background fields, most recent experiments additionally utilize the ability to reverse the orientation of the induced molecular electric dipole moment relative to the external bias field, and hence switch the sign of the EDM interaction. This technique is known as internal comagnetometry, and it requires the existence of suitable states, typically the $\Omega$-doublet of a $^3\Delta_1$ state~\cite{Leanhardt_2011, Loh_2013, ACME_2018, Roussy_2023}. A limitation of molecules which offer these states is that laser-cooling them is impractical, which makes it difficult to obtain large and cold samples. Therefore, it has been proposed to instead use $\ell$-doublets found in linear triatomic molecules or $K$-doublets found in symmetric-top molecules. They allow combining long lifetime with laser-coolable molecular structure, and can be polarized in low electric fields~\cite{Kozyryev_2017a, Hutzler_2020, Augenbraun_2021a,Mitra_2021,Vilas_2022,Lasner_2024}.

\begin{figure}
    \centering
    \includegraphics{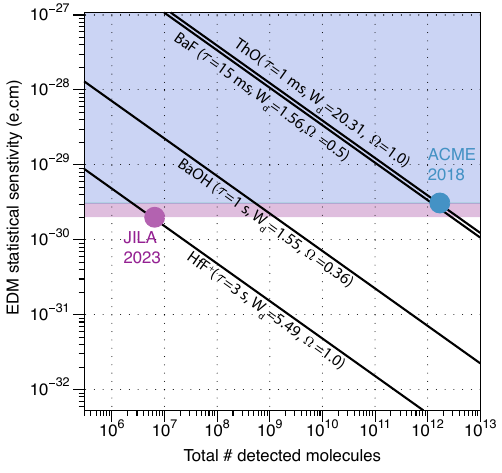}
    \caption{Overview of statistical sensitivity of different experimental platforms. Each line corresponds to a combination of $\tau$, $|\langle \Omega \rangle|$ and $W_\mathrm{d}$ [in units of $10^{24}$ $h\,$Hz/($e\,$cm)] which is typical for a given molecule species. The dots represent the two most recent experimental results~\cite{ACME_2018, Roussy_2023}, taking into account only statistical uncertainty. The blue and purple shaded regions are excluded by the ThO and HfF$^+$ experiments, respectively. 
    The $\langle \Omega \rangle$ value for not fully polarized BaOH is calculated in Section~\ref{efield-dependence-of-magnetic-moment}.
    It can be seen that our target sensitivity of $10^{-30}\,e\,$cm can be reached with $N_\mathrm{p} = 5 \times 10^9$ BaOH molecules at the shot noise limit.}
    \label{fig-sensitivity}
\end{figure}

Under the assumption of a shot-noise-limited measurement~\cite{Itano1993}, the EDM can be determined with a statistical error of
\begin{equation}
    \label{eq-sensitivity}
    \sigma = \frac{\hbar}{2 |\langle \Omega \rangle| W_\mathrm{d} \tau \sqrt{N_\mathrm{p}}},
\end{equation}
where $\tau$ is the free evolution time of the spin precession, $N_\mathrm{p}$ is the total number of detected particles and $\langle \Omega \rangle$ is the state sensitivity factor. Here, $\Omega = (\mathbf{J}_\mathrm{e}/\hbar) \cdot \mathbf{n}$ is the projection of the reduced total electronic angular momentum operator $\mathbf{J}_\mathrm{e}/\hbar$ onto the molecular axis, since in linear molecules this axis is aligned with the molecular-frame electric dipole moment, whose direction is given by the unit vector $\mathbf{n}$. Fig.\ \ref{fig-sensitivity} shows the result of this calculation for certain molecular species. For molecules in the $^2\Sigma$ electronic state, such as BaOH, $\langle \Omega \rangle \cong \langle \Sigma \rangle$ (strict equality holds only in the non-relativistic limit since the relativistic value of $\langle \Lambda \rangle$ is negligibly small but not zero), where $\Sigma$ is the projection of the reduced total electronic spin operator $\mathbf{S}_\mathrm{e}/\hbar$ onto the same axis. Hence, in this paper, we use $\langle \Sigma \rangle$ for the state EDM sensitivity factor and the product $\langle \Sigma \rangle \cdot W_\mathrm{d}$ for the overall EDM sensitivity. 
Note that the mean value of $\langle \Omega \rangle$ (or $\langle \Sigma \rangle$) is taken over the full molecular state in a particular electric field and this does not correspond to the pure electronic mean value in the molecular frame; as a result, this is not strictly $1/2$ for molecules in the $^2\Sigma$ electronic state which are not fully polarized.

To achieve the lowest possible bound on the eEDM, $W_\mathrm{d}$, $\langle \Sigma \rangle$, $\tau$ and $N_\mathrm{p}$ should all be large. Previous experiments made tradeoffs to reach either large $\tau$ or $N_\mathrm{p}$. For example, the ACME experiment is based on a cold molecular beam, which allows a large molecule number but limits $\tau$ to a few milliseconds~\cite{ACME_2018}. On the other hand, the HfF$^+$ experiment at JILA uses molecular ions in an rf trap, which is limited to trapping $\sim$20000 molecules at a time due to their strong electrostatic repulsion. Nonetheless, with a detection of $\approx120$ molecules in each shot and $\tau = \SI{3}{s}$, the JILA measurement reached the current world-leading statistical error of $\sigma = 2 \times 10^{-30}\,e\,\text{cm}$~\cite{Roussy_2023, Caldwell_2023}. In contrast, certain trapped heavy polyatomic molecules promise simultaneously achieving large values for all four parameters, $W_\mathrm{d}$, $\langle \Sigma \rangle$, $\tau$ and $N_\mathrm{p}$~\cite{Kozyryev_2017a}. The number of neutral molecules in an optical dipole trap can reach $10^5$ before density-dependent effects start becoming a problem, and they can be held for seconds. Currently, molecule sources and cooling become limiting factors before this point, but this is a rapidly advancing area~\cite{Lasner_2024}. 

Due to the presence of the heavy $^{138}$Ba atom in BaOH, its sensitivity is about $W_\mathrm{d} = 3.10\times 10^{24}\,h$~Hz/($e\,$cm)~\cite{Denis2019}. In the vibrational level with a single quantum of bending motion, labeled\footnote{The vibrational state labeling is $(\nu_1\nu_2\nu_3)$, where $\nu_i$ indicates the number of vibrational quanta in the Ba--O stretch, Ba--O--H bend, and O--H stretch modes respectively~\cite{Kozyryev_2017a}.} (010), this sensitivity can be reached at comparatively small external $E$-fields due to the small splitting between the opposite parity $\ell$-doublet levels (see Ref.~\cite{Jadbabaie_2023} for further explanation of the relevant properties of linear polyatomic molecules and their quantum numbers). While there are species with larger $W_\mathrm{d}$, these contain even heavier elements, which brings additional difficulties with cooling and trapping. We believe that Ba-containing molecules lie in the optimal regime where creating and cooling larger numbers of molecules is simple enough to compensate for the loss in sensitivity compared to species containing, for instance, Hg or Yb~\cite{Mitra_2021, Jadbabaie_2023}. A recent study even suggests that certain isotopologues of molecules of medium mass such as SrOH are actually very promising candidates for constraining the EDM parameter space~\cite{Gaul_2023}. Many of the results presented here also apply to other candidate species, such as SrOH, YbOH, and BaOCH$_3$.

\begin{figure*}
    \centering
    \includegraphics{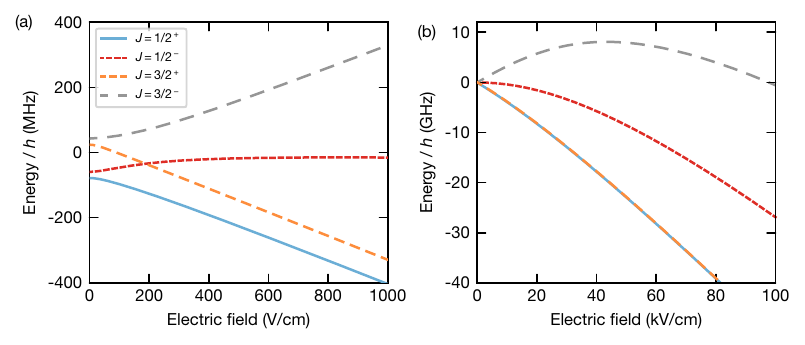}
    \caption{Simulation of Stark shift for certain states of BaOH. All shown states are in the rovibronic state $\tilde{X}{}^2\Sigma^+$(010), $N=1$ and in the hyperfine state $F=1, |m_F|=1$. We propose using $J=1/2^+, F=1, m_F=\pm1$ for the EDM measurement. Focusing and deceleration will be done in $J=3/2^-$ instead, due to its low-field-seeking character at intermediate fields. Both subfigures show the same result, with (a) focusing on the low-field region where the structure is dominated by the $\ell$-splitting, and (b) focusing on the high-field region where the rotational structure dominates. The colors correspond to the same states in both subfigures. In (b), the states $J=1/2^+$ and $J=3/2^+$ overlap and can't be distinguished. The calculations are based on the model by Kozyryev and Hutzler~\cite{Kozyryev_2017a}.}
    \label{fig-stark-curves}
\end{figure*}

\subsection{Noise}\label{sec-noise}
We have analyzed the requirements to reach a statistical EDM sensitivity of $10^{-30}\,e\,\text{cm}$, two times below the current state of the art. Using BaOH with a coherence time of \SI{1}{\second}, this goal requires measuring a change of \SI{1}{\micro\hertz} on the spin-precession frequency. The measurement of a single molecule has quantum projection noise of 80~mHz from Eq.~\ref{eq-sensitivity}, so our goal requires averaging $5\times 10^9$ measurements while ensuring no other noise sources add excess noise or systematic errors. With such a stringent requirement, it is highly important to carefully quantify noise in the experiment. In this section we briefly introduce the role of noise in general terms, in preparation of the subsequent sections in which we discuss noise sources specific to the proposed approach.

How detrimental noise is depends on its frequency. The most problematic contributions are close to the repetition rate, i.e., $f \approx 1/\tau$. Homogeneous noise (i.e., noise which affects all molecules equally) at this frequency will directly degrade the performance of the experiment because different frequency shifts will be measured in each repetition. Inhomogeneous noise will instead decrease the spin-precession contrast, with a similar resulting loss in sensitivity. In either case, if the noise is too strong, the experiment can't reach the shot-noise limit in Eq.~\ref{eq-sensitivity}. At higher frequencies, the effect of noise will often be suppressed as $1/f$. Noise at frequencies below $1/\tau$ is also suppressed, proportional to $f$, because each experimental run measures a difference of the spin-precession frequency of two samples, removing the dependence on absolute stability. Finally, there can be noise which is correlated with the direction of the effective $E$ field experienced by the valence electrons whose EDM we probe. Such effects are extremely problematic because they do not simply reduce the measurement precision, but instead look like an EDM signal. Fortunately, the use of internal comagnetometry suppresses most possible sources of correlated noise, though there are exceptions, as discussed for example in Refs.~\cite{Caldwell_2023, Anderegg_2023}.

Using a trap is both a blessing and a curse when it comes to noise -- on the one hand, confining the molecules to a small volume reduces the effect of inhomogeneity of the dc background fields as well as that of the motional magnetic field; on the other hand, the trap can also add noise, especially near the trap frequency, caused by the differential energy shift which molecules experience as they move through the trap potential (see Section \ref{sec-optical-trapping}).

\subsection{Field stability}
\label{sec-field-stability}
Because of the electron's large magnetic dipole moment, a crucial aspect of an EDM experiment is careful control of the magnetic field. This can be simplified by reducing the effective molecular magnetic moment $\mu$, which determines the Zeeman shift of the EDM-sensitive state pair via $\Delta E_Z = \mu B m_F$. In current experiments with ThO and HfF$^+$, this is done by choosing a $^3\Delta_1$ state, where the magnetic moments associated with the electronic spin and orbital angular momentum cancel out to roughly 1\%~\cite{ACME_2018,Roussy_2023}.

With polyatomic molecules, there is another way to do this since $\mu$ depends on the applied electric field~\cite{Kozyryev_2017a,Anderegg_2023}. Calculations presented in Section \ref{efield-dependence-of-magnetic-moment} show that the state pair $N=1, J=1/2^+, F=1, m_F\pm1$ exhibits a zero-crossing of $\mu$ at $E=$~\SI{52.3}{V/cm}. At this field, the molecule is approximately 25\% polarized, with Stark shifts as shown in Fig.~\ref{fig-stark-curves}. A tunable $g$-factor means that, with careful control of a relatively small applied $E$ field, the magnetic moment can be significantly reduced. Then, the optimal values for $E$ and $B$ depend on the stability of the $E$ field and the slope of $\mu(E)$ around its zero-crossing. Specifically, the rms frequency error caused by field fluctuations is given by
\begin{equation}
\label{eq-field-errors}
    \sigma = \frac{1}{h}\sqrt{\left(\frac{\partial \mu}{\partial E} \sigma_E B \right)^2 + \left(\mu \sigma_B\right)^2},
\end{equation}
where $\sigma_E$ and $\sigma_B$ denote the rms error of $E$ and $B$ in the relevant frequency range and $\sigma$ is the error budget for field-dependence, which we assume here to be \SI{1}{\micro\hertz}. The first term can be reduced by lowering the bias magnetic field, while the second term can be reduced by operating near the $g$-factor zero-crossing. 

Typically, a bias magnetic field is used to ensure maximal sensitivity to the precession frequency by applying a $n\pi + \pi/2$ offset to the free evolution phase. A subsequent measurement that projects the time-evolved state back onto the initial superposition state is therefore linearly sensitive to the EDM. However, in the case of a vanishing bias field, the measurement sensitivity can instead be maximized by rotating the readout basis by $\pi/2$ relative to the initial superposition basis. This readout rotation also achieves linear sensitivity to the EDM but allows reducing the bias magnetic field and thereby the noise caused by $\sigma_E$.

If the experiment is operated near the $g$-factor zero-crossing, the magnetic noise contribution from $\sigma_B$ can also be significantly suppressed. The $g$-factor slope near the zero-crossing is given by $|\partial \mu / \partial E| = h \times \SI{250}{\milli\hertz\per\nano\tesla\per(\volt\per\centi\meter)} = 1.8\times 10^{-2}$~$\mu_B/(\text{V/cm})$. We can parameterize the magnetic moment as $\mu = \partial \mu / \partial E \times \delta E$, where $\delta E = E-E_0$ is the detuning of the bias electric field from the zero-crossing at $E_0$.

Tuning the $g$-factor \textit{exactly} to zero is not possible since the bias magnetic field also serves to define a quantization axis for the electron spin. Without any bias field, the $m_F=\pm 1$ levels are degenerate, and any magnetic fields transverse to the bias electric field will couple the states at second order in perturbation theory, reducing spin precession contrast. The application of a bias magnetic field lifts this degeneracy and protects against the transverse field noise. For the states considered here, the quadratic transverse magnetic sensitivity~\cite{Anderegg_2023} is calculated to be $\sim 4~\text{\textmu Hz}/\text{nT}^2$. To suppress sensitivity to transverse fields generated by the magnetic noise $\sigma_B$, we require $\mu B \gg h\times4~\text{\textmu Hz}/\text{nT}^2 \times \sigma_B^2$. We see that for reasonable magnetic field control, the bias magnetic field can be reduced significantly before running into the issue of transverse field sensitivity. 

\begin{figure}
    \centering
    \includegraphics[width=\columnwidth]{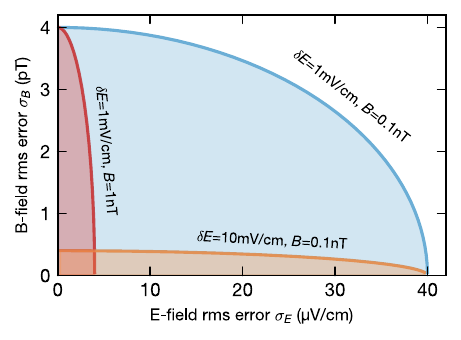}
    \caption{Noise requirement to reach a total contribution to the error of the spin-precession frequency of \SI{1}{\micro\hertz} at different combinations of the electric and magnetic background fields. The shaded regions indicate the range where the errors are sufficiently low. The curves are calculated for $(\partial \mu / \partial E)/ h =\SI{250}{\milli\hertz\per\nano\tesla\per(\volt\per\centi\meter)}$, the value for the level $N=1, J=1/2^+, F=1$.}
    \label{fig-sigmaE-vs-sigmaB}
\end{figure}

To achieve the target error budget of $1~\text{\textmu Hz}$, we require each noise source in Eq.~\ref{eq-field-errors} to be at or below this level. For the states considered here, we obtain the following limit on electric and magnetic noise: $\sigma_E B \sim \sigma_B \delta E < 4\times 10^{-6} ~\text{nT} \times \text{V/cm}$. For example, operating at $\delta E = 10~\text{mV/cm}$, which results in $\mu = 1.8 \times 10^{-4}~\mu_B$, requires $\sigma_B < 400~\text{fT}$. Similarly, if we operate at $B=0.1~\text{nT}$, we require $\sigma_E<40~\text{\textmu V/cm}$. In Fig.\ \ref{fig-sigmaE-vs-sigmaB}, we include a plot of various operating conditions that achieve the target noise performance. Achieving this target is critical for field noise occurring at frequencies near the experiment cycle rate. For higher frequency noise, averaging over many experimental runs further reduces the sensitivity to field noise.

In any case, the requirements on field stability demand careful experiment design, which we explain in Section \ref{sec-implementation-fields}. This task would become easier with smaller $\partial \mu / \partial E$. For example, in the case of BaOH it is reduced when using the $N=2$ rotational level compared to $N=1$. However, this is at the cost of having to work at a larger $E$ field and a reduced $\langle \Sigma \rangle$ (see Sec.\ \ref{efield-dependence-of-magnetic-moment}). Whether this is an improvement in an actual experiment will depend on whether the dominant noise contribution is constant or proportional to the applied field.

\subsection{Optical trapping}
\label{sec-optical-trapping}
The key ingredient in this proposal is optical dipole trapping. It makes use of the ac Stark effect, which lowers the energy of certain molecular states in a light field~\cite{Grimm_2000}. Neutral molecules can also be trapped in a number of other ways, but significant development of the experimental techniques is necessary before these can be used in eEDM experiments. For example, microwave traps are theoretically possible, but have never been successfully demonstrated with molecules~\cite{Wright_2019}, and electrostatically trapped molecules are expected to have their spin precession disturbed by accumulating a geometric phase~\cite{Tarbutt_2009}.

Optical dipole traps, on the other hand, are well-suited for our task because the energy shifts they create can be understood and controlled to an extremely high level of precision, as proven by optical lattice clocks~\cite{Derevianko_2011, Abdel-Hafiz_2019}. They are also an established technology which works generically for almost any molecule, as long as suitable lasers are available. A typical choice is \SI{1064}{nm}, where high-power lasers are relatively affordable, and which is far red-detuned to the strong electronic transitions of many interesting molecules.

In order to trap a molecule the trap has to shift its energy levels, and this shift is in general dependent on the state. At a given temperature $T$, the trap shift needs to be on the order of $10\, k_B T$ to avoid evaporation, with the Boltzmann constant $k_B$. This means that at typical temperatures of \SI{50}{\micro\kelvin}~\cite{Hallas_2024}, a shift of $h \times \SI{10}{MHz}$ is required. Meanwhile, the energy difference between the EDM-sensitive state pair needs to be sufficiently controlled to achieve \textmu Hz overall precision. Because trap shifts are inhomogeneous, the trap noise contribution can be reduced by averaging over multiple molecules. Therefore, our requirement is that the trap noise must be less than the single molecule projection noise of 80~mHz, obtained from Eq.~\ref{eq-sensitivity}. With the absolute shift at $\SI{10}{MHz}$, the differential shift must be at least 8 orders of magnitude suppressed. This is only possible with a detailed understanding of the physics involved. 

In general, the effective Hamiltonian for the interaction of the ac polarizability $\alpha(\omega)$ of an atom or molecule with a light field of frequency $\omega$, polarization $\hat{\varepsilon}$, and amplitude $E_0$, is given by~\cite{Kien_2013, Caldwell_2020a}
\begin{multline}
    H = -\frac{E_0^2}{4} \sum_{k=0}^2 T^k\left(\alpha(\omega)\right) \cdot T^k(\hat{\varepsilon},\hat{\varepsilon}^*)\\ 
    = -\frac{E_0^2}{4} \sum_{k=0}^2 \sum_p (-1)^p T^k_p (\alpha(\omega)) \, T^k_{-p} (\hat{\varepsilon},\hat{\varepsilon}^*)\\
    = -\frac{E_0^2}{4} \sum_{k=0}^2 \sum_{p,q} (-1)^p \mathcal{D}^{(k)}_{pq}(\Omega)^* \, T^k_q (\alpha (\omega)) \, T^k_{-p} (\hat{\varepsilon},\hat{\varepsilon}^*),
\end{multline}
where we have written the interaction in spherical tensor form, summing over the tensor rank $k$ and tensor components $p$ and $q$. The index $p$ labels the lab-frame projections and $q$ labels molecule frame projections, with $|p|,|q| \leq k$. The quantization axis in the lab-frame (corresponding to $p=0$) is taken to be along the lab-frame $\hat{X}$-axis defining $m_F$, which is also taken to be along the applied magnetic field $B$. The quantization axis in the molecule-frame (corresponding to $q=0$) is taken to lie along the internuclear axis vector $\mathbf{n}$. The Wigner rotation matrix element $\mathcal{D}^{(k)}_{pq}(\Omega)^*$ transforms from lab-frame component $p$ to molecule-frame component $q$ according to the Euler angles labeled by $\Omega$.

The polarizability operator can be thought of as a detuned two-photon coupling. The $k=0, 1$, and $2$ terms correspond to scalar, vector, and tensor polarizabilities, respectively, and can be expressed in terms of the Cartesian components of $\alpha$ in the molecule frame. For linear molecules with cylindrical symmetry, the only non-zero molecule-frame components are $\alpha_{xx}, \alpha_{yy}, \alpha_{zz}$. For an electronic state of pure $\Sigma^+$ symmetry, the polarizability is isotropic perpendicular to $\mathbf{n}$, and as a result $\alpha_{xx} = \alpha_{yy} := \alpha_\perp$, $\alpha_{zz} := \alpha_\parallel$, and only the $q=0$ components of $T^k_q(\alpha)$ are non-zero. However, we note that vibronic and spin-orbit couplings can distort cylindrical symmetry and admix states with other electronic symmetries, resulting in $q=\pm2$ terms in the effective Hamiltonian~\cite{Jadbabaie_2023}. Even if the ground state is unperturbed, these perturbations occur in excited $\Pi$ states and can couple to the ground state via the off-resonant trapping light, though such mixings are expected to be small~\cite{Zhang_2023a}. Fortunately, such $q=\pm 2$ terms vanish for differential measurements using states with equal values of $\langle \mathbf{n} \rangle$, i.e. equal molecular orientations, which includes the T-reverse $m_F =\pm 1$ states we consider. Therefore, for the remainder of the discussion, we consider only the $q=0$ terms.

Examination of both the polarizability and polarization tensors indicates the scalar $k=0$ term is proportional to the trace of $\alpha$ and proportional to $E^2$, making it independent of both $\hat{\varepsilon}$ and $m_F$. Further, the differential scalar shift is suppressed by the ratio of Zeeman splitting of the EDM states to the laser detuning, a suppression factor of at least 14 orders of magnitude. The $k=1$ vector terms of the polarization tensor are antisymmetric and vanish for linear polarization, and mimic the effects of magnetic fields. Finally, the $k=2$ terms are symmetric and traceless, and mimic the effects of electric fields.

The polarizability shifts can be managed and surpressed by careful control of the lab-frame components of the polarization tensor $T^k_p(\varepsilon,\varepsilon^*)$. Both the vector and tensor components generically contain $p\neq 0$ components that can mix states with $\Delta m_F \neq 0$. However, these off-diagonal components can be suppressed by a variety of mechanisms, such as large sublevel splittings. In the case of measurements considered here, Stark splitting of neighboring $m_F$ states suppresses contributions from $p=\pm1$ terms. The $k=1, p=0$ vector component can mimic parallel magnetic fields and must be suppressed by using linearly polarized light and choosing the wavevector of the trapping light to be perpendicular to the quantization axis. Creating light fields with purely linear polarization is difficult, but circular polarization of $10^{-8}$ has previously been experimentally demonstrated~\cite{Zhu_2013a}. In addition, $k$ can be chosen such that $(\hat{k} \cdot \hat{B}) \approx 10^{-3}$, where quantities with hats are unit vectors. Further, the effect of vector shifts are further suppressed by the use of first-order magnetically insensitive states (that is with $\langle M_S\rangle \approx 0$). Nonetheless, the vector light shift remains a concern and must be carefully characterized. If it poses a problem despite the best efforts to create pure linearly polarized light, another possibility is to go to larger laser detuning, where the relative contribution of the vector component becomes smaller \cite{Bonin1997, Caldwell_2020a}.

The tensor components must be considered for linear polarization. Due to the structure of the electric field polarization tensor, the $k=2, p=\pm1$ and $k=2, p=\pm2$ tensor terms vanish in the limit of $\hat{\varepsilon} \parallel \hat{X}$, that is linearly polarized trapping light along the $B$ field quantization axis. However, the $p=\pm2$ terms can directly couple the two $m_F=\pm 1$ states and reduce measurement contrast, and must be considered in detail. The spherical tensor form of this interaction with the light field is scaled by $\sin^2{\theta}$, where $\theta$ is the light polarization angle with the $B$ field quantization axis. For $\theta\sim 10^{-3}$, the suppression is $\sim 10^{-6}$. Additionally, the tensor shift is further suppressed compared to the scalar shift by approximately an additional order of magnitude due to a combination of the state's $J=1/2$ content (which does not have strong tensor shifts given $|J|<1$) and the slight reduction in magnitude of the tensor polarizability reduced matrix element $T^2_{q=0}(\alpha)$ compared to the scalar component $T^0_{q=0} (\alpha)$. The combination results in a $T^2_{p=\pm2}(\alpha)$ coupling on the order of $\sim 100$~mHz. To further mitigate this, the Gaussian field in the cavity could be designed to have minimum curvature, with the shape of the magnetic field matched to maintain perpendicular conditions as much as possible. The remaining $k=2, p=0$ diagonal tensor term is independent of the sign of $m_F$, but can mimic electric fields, causing the $g$-factor of the EDM state to depend on the trapping light intensity.

In an earlier characterization of zero $g$-factor states in CaOH $\tilde{X}{}^2\Sigma^+ (010)$~\cite{Anderegg_2023}, it was found that the zero crossing of the $g$-factor varied as the molecules sampled the intensity distribution of the trapping potential for linearly polarized light parallel to the $B$ field. This results in precession noise at the trap oscillation frequency. For BaOH, using the calculated dynamic polarizabilities in Figure~\ref{fig-dyn-pol} (discussed in more detail in Section~\ref{sec:calculations}), we compute a $\sim$236~mV/cm shift in the $g$-factor zero crossing, resulting from 10\% variation of the trap intensity required to obtain $h \times \SI{10}{MHz}$ scalar trapping shift. This gives a $g$-factor variation of $h\times$60~mHz/nT~$\approx 4.3\times 10^{-3}~\mu_B$, which translates to 6~mHz for the $B$ fields we consider. With at least $4\times10^7$ measurements, the averaged tensor shift noise contributes $<1$~\textmu Hz noise. 

Since the EDM signal is symmetry-violating, a vector light shift by itself cannot cause a false EDM; however, a vector light shift could couple to imperfections which can mimic symmetry-violating effects, such as stray electric fields~\cite{Baron2017ACMELong,Anderegg_2018}. Systematics arising from these shifts are expected to be smaller than the proposed sensitivity~\cite{Anderegg_2018}, and could be both quantified and further suppressed by operating at an electric field where the EDM sensitivity vanishes or changes sign~\cite{Kozyryev_2017a,Anderegg_2023}. 

The proposed approach is different from the use of a magic wavelength. There, the differential polarizability between different rovibronic levels is tuned to zero~\cite{Katori_2003, Bause_2020, Leung_2023} 
due to crossing polarizability curves 
with different slopes at a particular trapping wavelength. 
In our case, the curves of the two $m_F$ levels are automatically practically parallel at \SI{1064}{nm}. 

Because the two states that form the superposition only differ in $m_F$, we will very likely also not be limited by higher-order polarizabilities caused by electric quadrupole or magnetic dipole transitions, or by second-order hyperpolarizability. These can all be problematic for optical lattice clocks, but their differential components are extremely small in our case.

\subsection{Molecule number and lifetime}
\label{sec-lifetime}
To achieve an EDM measurement beyond the current best limit with BaOH at $\tau =\SI{1}{s}$, the total molecule number $N_\mathrm{p}$ should be at least $5 \times 10^{9}$. Assuming a cycle time roughly equal to $\tau$ and a total run time of 120 days, this requires approximately 500 molecules to be detected in a single run. With a $1/\mathrm{e}$ lifetime of \SI{0.86}{s}, a transport efficiency of 50\%, and a near-unity detection efficiency, the number initially loaded into the optical trap should be around 2600. This is a realistic goal, considering that the highest reported number of trapped polyatomic molecules is 2900, obtained with CaOH~\cite{Hallas_2024}. Possible strategies for improving this result include more efficient MOT capture, better cooling and using a larger and deeper dipole trap.

A number of factors can cause molecule loss during the free-evolution time. The dominant one is the spontaneous decay of the EDM-sensitive state (010) into the vibrational ground state. Based on a pseudo-relativistic electronic-structure calculation using density functional theory, the lifetime is \SI{0.86(9)}{s}, similar to the case of CaOH~\cite{Vilas_2023}. 

Molecules can also collide with each other, undergoing chemical reactions. Calculating the rate constant for this process from first principles is notoriously difficult, but luckily there are now experimental measurements for the cases of CaF, SrF, and CaOH which found rates of $2\times 10^{-10}$ to \SI{4e{-10}}{\cubic\centi\meter \per\second}~\cite{Anderegg_2021, Jorapur_2023, Vilas_2025}. If BaOH behaves similarly, we can work at a reasonable density of \SI{e{9}}{\per\cubic\centi\meter} without being limited by collisions. The real loss coefficient may be somewhat larger due to attractive head-to-tail collisions of polarized molecules~\cite{Ni_2010a}, but even then, collisional loss is unlikely to be the dominant factor in our experiment. 

The lifetime can also be limited by interaction with the electromagnetic field, both from the trap and from blackbody radiation~\cite{Hoekstra_2007}. The photon-scattering rate of a dipole trap can be estimated for the case of a molecule with a single dipole transition to be proportional to the trap depth~\cite{Grimm_2000}:
\begin{equation}
    \gamma \approx \frac{U}{\hbar} \frac{\Gamma}{\Delta},
\end{equation}
where $\gamma$ is the scattering rate, $U$ is the trap depth, and $\Gamma$ and $\Delta$ are the transition linewidth and the detuning from the transition, respectively. This indicates that for BaOH with the lowest dipole-allowed electronic transition at \SI{871}{nm} in a 1064-nm trap we have $\hbar\gamma/U \approx 5\times10^{-8}$, or equivalently, that the lifetime in a 10-MHz deep trap is \SI{2}{s}, if we assume a linewidth of \SI{3}{MHz}. To reach longer lifetimes, one can either work at a larger detuning or at a lower temperature, which allows using a lower trap depth. 

In contrast to this, room-temperature blackbody radiation couples predominantly to rovibrational transitions, which can also cause effective loss of molecules. For the case of CaOH, the blackbody-radiation-limited lifetime is about \SI{1}{s} at room temperature~\cite{Vilas_2023}, and it is expected to be similar for BaOH, extrapolating from the results in Ref.~\cite{Buhmann_2008}. The blackbody-radiation-induced loss can probably be reduced by a factor 5 by cooling the surroundings of the molecules to \SI{77}{K}. 

Finally, there are background-gas collisions, which are unlikely to be a problem because lifetimes of atomic gases can reach \SI{100}{s} with background pressures on the order of \SI{e{-11}}{mbar}~\cite{Park_2022}.

\section{Calculations}\label{sec:calculations}
\subsection{Electric dipole polarizability}

\begin{figure}
    \centering
    \includegraphics[width=\columnwidth]{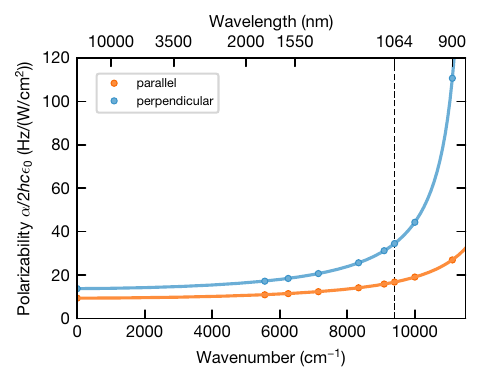}
    \caption{Dependence of the parallel and perpendicular components of the dynamical polarizability $\alpha/(2hc\epsilon_0)$ (in intensity-based units) on the laser frequency. The vertical dashed line represents the considered 1064~nm laser frequency.}
    \label{fig-dyn-pol}
\end{figure}

Planning a dipole trap for this experiment requires a quantitative understanding of the molecule's electric dipole polarizability. The static limit of the scalar polarizability $\alpha^S(0)$ was calculated using the relativistic coupled cluster (CCSD(T)) method implemented in the DIRAC package~\cite{dirac-paper,DIRAC19} by means of the finite-field perturbation approach. Augmented valence Dyall basis sets~\cite{Dyall2009,Dyall2016} were used in the calculations together with the extrapolation to the complete basis set limit. The internuclear bond lengths were extracted from experimental data~\cite{Kinsey86}. Various computational parameters (basis set composition, extrapolation schemes, active space, relativistic Hamiltonians) were thoroughly investigated and the associated uncertainties were assigned. The predicted value of the static scalar (isotropic) polarizability is  $\alpha^S(0)=~(65\pm3)~h\,\text{mHz}/(\text{V}/\text{cm})^2$. For comparison with the dynamical polarizabilities in Fig.~\ref{fig-dyn-pol} this is expressed in terms of the intensity as $\alpha^S(0)/(2hc\epsilon_0) = (12.3\pm0.6)~\text{Hz}/(\text{W}/\text{cm}^2)$. 

The dynamical laser frequency dependence of the scalar polarizability $\alpha^S(\omega)$ was calculated at the coupled cluster (CCSD) level of theory in the pseudo-relativistic framework based on effective core potentials implemented in the CFOUR~\cite{cfour,cfour-paper} program package employing linear response theory. To this effect, we employed Dunning's pseudopotential basis sets~\cite{Kendall1992,Lim2006,Hill2017}. Similarly to the static limit, various computational parameters were converged out and the corresponding uncertainties accounted for, as detailed in the upcoming theoretical publication~\cite{polar-theor}. The predicted value of the scalar polarizability is $\alpha^S(1064\,\text{nm})/(2hc\epsilon_0) =(28.6\pm2.3)~\text{Hz}/(\text{W}/\text{cm}^2)$.

Using our computed value for the perpendicular polarizability $\alpha_\perp$ and the available transition frequencies to the nearest electronic $\Pi$ state components $\omega_{1/2}$ and $\omega_{3/2}$ \cite{Wang2008}, we can calculate an estimate of the vector polarizability using an approximate relation (neglecting contributions from the higher states)~\cite{Caldwell_2020a}
\begin{equation}
\alpha^V=\frac{\omega}{2} \frac{\omega_{3/2}-\omega_{1/2}}{\omega_{1/2}\omega_{3/2}} \alpha_\perp
\end{equation} giving 
$\alpha^V(1064\,\text{nm})/(2hc\epsilon_0) = (6.6\pm0.4)~\text{Hz}/(\text{W}/\text{cm}^2)$.

\subsection{Electric-field dependence of molecular magnetic moment and state EDM sensitivity}
\label{efield-dependence-of-magnetic-moment}

\begin{figure}
    \centering
    \includegraphics[width=\columnwidth]{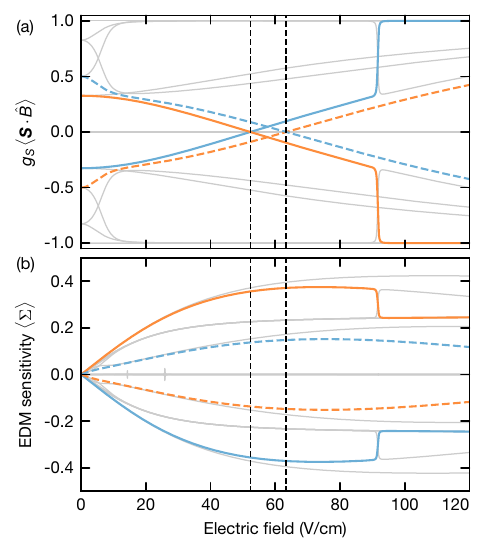}
    \caption{Dependence of (a) the electron spin magnetic moment $ g_S \, \langle \bm{S}\cdot\hat{B} \rangle$, with $\hat{B}$ being the unit vector in the direction of the applied magnetic field), and (b) the state EDM sensitivity $\langle \Sigma \rangle$, on the electric field. The solid and dashed lines denote the $|N=1, J=1/2^+,\,F=1,\,m_F=\pm1\rangle$ and $|N=1, J=3/2^+,\,F=2,\,m_F=\pm1\rangle$ state pairs, respectively, and the blue and orange lines denote $m_F= +1$ and $m_F= -1$ states. The gray curves represent other states in the $N=1$ manifold.The vertical dashed lines indicate the position of the $g$-factor zero-crossings.}
    \label{fig-gfactor-vs-E-N1}
\end{figure}

The sensitivity of the magnetic moment of BaOH to an applied electric field has been analyzed using the same effective Hamiltonian and methodology that was recently employed to study the CaOH molecule~\cite{Anderegg_2023}. It is assumed that this molecular magnetic moment is dominated by the electron spin magnetic moment $\bm{\mu}_S=-g_S\mu_B\bm{S}$, where $g_S$, $\mu_B$, and $\bm{S}$ are the electron spin $g$-factor, the Bohr magneton, and the total reduced electron spin angular momentum, respectively.

All the field-free Hamiltonian parameters were taken from Ref.~\cite{Fletcher1995}, except for two: the molecular-frame electric dipole moment, which was extracted from Ref.~\cite{Frey2011}, and the hyperfine structure parameters associated with the hydrogen nucleus ($b_F=-0.70$~MHz, $c=1.00$~MHz). These hyperfine structure constants were calculated using the DIRAC code, employing the Dirac--Coulomb Hamiltonian and the finite-field method. The dyall.cv3z basis set was used,
and the experimental internuclear bond lengths were extracted from Ref.~\citenum{Kinsey86}.
The electron correlation effects were taken into account through the application of the Fock-space coupled-cluster method (correlating molecular orbitals with energies between $-800~E_\mathrm{h}$ and $800~E_\mathrm{h}$, $E_\mathrm{h}$ being the Hartree energy). All the calculations were performed in the absence of the trap. Based on the CaOH case~\cite{Anderegg_2023}, the changes to the zero $g$-factor crossings from including the trap interactions are of the order of 1 V/cm.


In Fig.~\ref{fig-gfactor-vs-E-N1} we show the electric field dependence of the effective molecular $g$-factor in the $\tilde{X}$(010) states of BaOH (only rotational states with $N=1$ are displayed). We found that for this system the molecular $g$-factor sensitivity to an applied electric field is comparable to that of CaOH. Zero-crossings with a smaller slope can be found at $N=2$, but at higher electric fields and reduced state EDM sensitivity. Given that the electron spin magnetic moment dominates the total magnetic moment of the molecule, the effective molecular $g$-factor can be expressed as $g_\text{eff}=g_S \, \mu_B \, \langle \bm{S}\cdot\hat{\bm{B}} \rangle$.
In Table~\ref{table:eEDM-sensitivity} we summarize the results of these calculations, including the derivative at the zero-crossing, $ \partial g_\text{eff}/\partial E|_{g_\text{eff}=0}$.

\begin{table}[tbh]
\renewcommand{\arraystretch}{1.25}
\caption{Electric field strengths at zero effective $g$-factor crossings for selected rotational states in the $N=1$ and $N=2$ manifolds, as well as $g_\text{eff}$ vs $E$ slopes and EDM state sensitivities $\langle \Sigma \rangle$ at these applied electric fields. We labeled the states in terms of their correlated zero-field quantum numbers $|N$, $J^p$, $F$, $m_F \rangle$}
\centering
\begin{tabular*}{\linewidth}{@{\extracolsep{\fill}} l rcr}
\hline
State & \multicolumn{1}{c}{$E$} & \multicolumn{1}{c}{$(1/h)\,\partial g_\text{eff}/\partial E$} & \multicolumn{1}{c}{$\langle \Sigma \rangle$} \\
 & \multicolumn{1}{c}{[V/cm]} & \multicolumn{1}{c}{$\left[\frac{\text{mHz/nT}}{\text{V/cm}}\right]$} &  \\[1ex]
\hline
$|1, 1/2^+, 1, \pm1 \rangle$ & 52.3  & $\pm250$  & $\mp0.36$ \\[1ex]
$|1, 3/2^+, 2, \pm1 \rangle$ & 63.4  & $\mp229$  & $\pm0.15$ \\[1ex]
$|2, 3/2^-, 1, \pm1 \rangle$ & 331.2 & $\pm145$  & $\mp0.15$ \\[1ex] 
$|2, 5/2^-, 3, \pm1 \rangle$ & 386.7 & $\mp125$  & $\pm0.09$ \\ 
\hline
\end{tabular*}
\label{table:eEDM-sensitivity}
\end{table}

\section{Implementation}\label{sec-implementation}
\subsection{Overview}
We envision a setup where molecules are created in a cryogenic buffer gas beam, then slowed in a Stark decelerator to the capture velocity of a magneto-optical trap (MOT). In this trap, molecules are held at rest relative to the lab frame, and cooled down enough to be trapped in an optical transport trap. This happens outside the magnetically-shielded volume where the EDM measurement is done, because MOTs require large $B$-field gradients which would compromise the shielding. The transport trap then vertically moves the molecules into the shielded volume, and deposits them into an optical lattice (the ``science lattice'') formed inside a buildup cavity, where Helmholtz coils and plate electrodes provide controlled static $B$ and $E$ fields. This is where the molecules are excited into the EDM-sensitive state and undergo spin precession. After the free evolution time, the spin-precession phase is read out optically, and the next repetition of the experiment can begin.

\begin{figure*}
    \centering
    \includegraphics{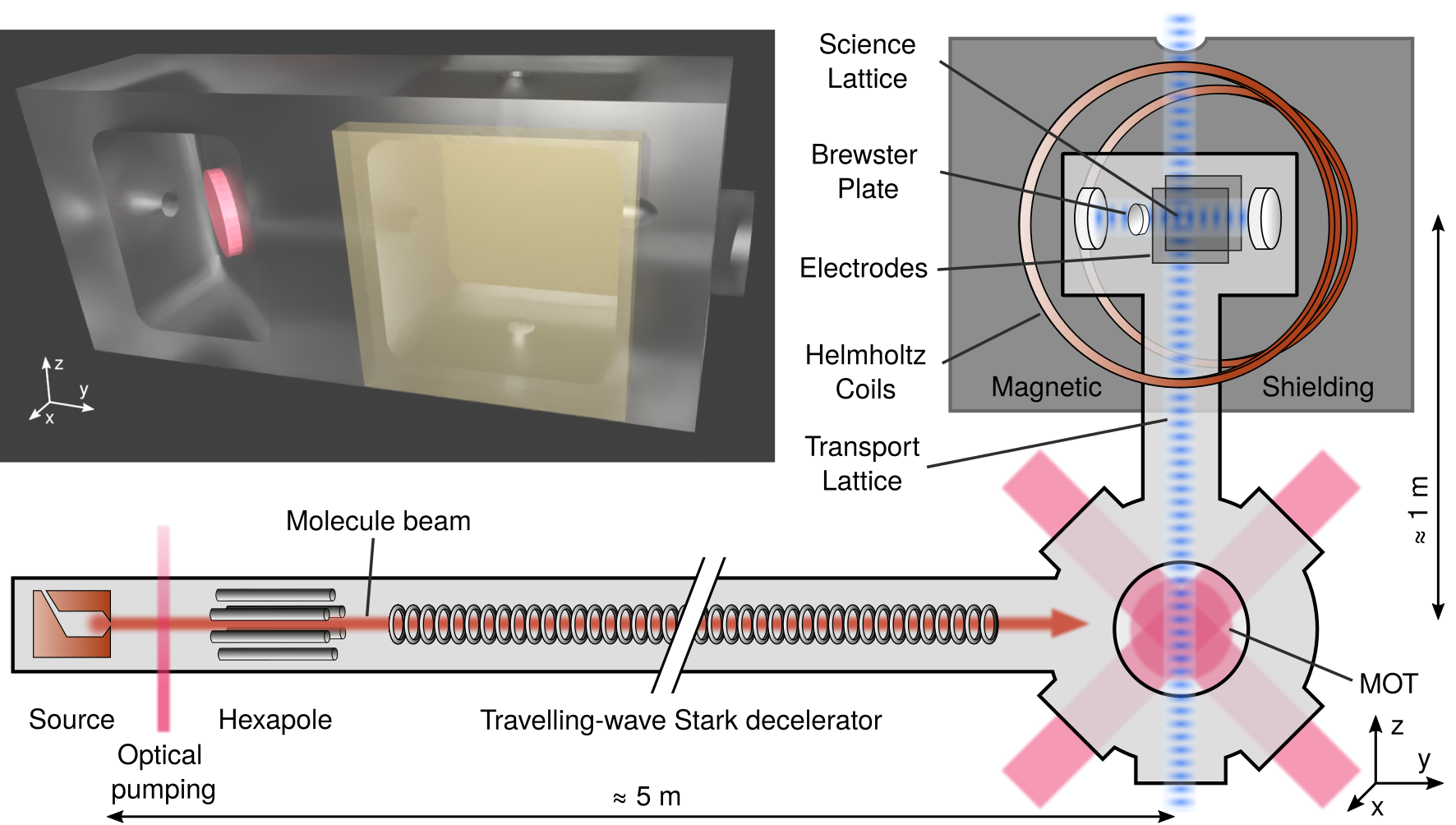}
    \caption{Sketch of the envisioned experimental apparatus. BaOH molecules are created in the source and fly along the brown arrow. They are collimated and slowed by the hexapole and decelerator. Additional hexapoles or transverse-laser-cooling stages may be added either before or after the decelerator, to avoid loss caused by beam divergence. Molecules are then slowed down to a standstill and cooled in the MOT, transported to the science lattice, and undergo spin precession there. The detection (not shown) is along the $X$-direction, through the transparent electrodes. Inset: Concept design of the glass science cavity assembly. The red element is a Brewster plate for polarization cleaning. The mirrors on the left and right side form the trapping cavity. The square glass pieces on the front and back are coated with ITO electrodes to create highly homogeneous $E$-fields. The spacer's size is \SI{10}{cm} along its longest direction.}
    \label{fig-setup}
\end{figure*}

\subsection{Molecule source}\label{sec-source}
During recent years, the cryogenic buffer-gas beam has been established as the method of choice for creating bright and slow molecular beams~\cite{Hutzler_2012, Truppe_2017a, Anderegg_2023}. We favour a design where barium is ablated from a metallic target. BaOH molecules are formed by injecting water or methanol into the source chamber where they react with the barium~\cite{Steimle_2019, Augenbraun_2021}. Molecules subsequently thermalize with a background of noble-gas atoms at cryogenic temperature. Under good conditions, a neon-cooled source can produce pulses of $10^{11}$ molecules per steradian~\cite{Mooij_2024}. When coming out of the source, the molecules are distributed over multiple rotational and vibrational levels with probabilities depending on the initial state of the reactants and the buffer-gas temperature~\cite{Jadbabaie_2020}.

The primary challenge with cryogenic sources is maintenance: in current experiments, they can usually only be run continuously for a few days until unwanted side-products of the chemical reactions, frozen buffer gas or reaction educts start causing trouble. The advantage of our setup is that the MOT only needs to be loaded once per second with a single pulse, such that the source can be operated at a small repetition rate. Still, continued development of better source designs which require less maintenance is a high priority to ensure undisturbed long-term data taking.

\begin{table}[tbh]
    \renewcommand{\arraystretch}{1.25}
    \centering
    \caption{Parameters of the envisioned experiment.}
    \begin{tabular}{l l}
        \hline
        Projected statistical sensitivity & $10^{-30}\,e\,\text{cm}$ \\
        Molecule number loaded into transport & 2600 \\
        Molecule number detected per run & 500 \\
        Total molecule number detected $(N_\mathrm{p})$ & $5 \times 10^9$ \\
        Molecule temperature & \SI{50}{\micro\kelvin} \\
        Spin-precession time $(\tau)$ & \SI{1}{s} \\
        Cycle time & \SI{1.2}{s} \\
        Total measurement time & $\approx \SI{120}{days}$ \\
        Applied $B$-field & \SI{0.1}{nT} \\
        Applied $E$-field & $ \SI{52.3}{V/cm}$ \\
        Science cavity length & \SI{10}{cm} \\
        Science cavity finesse & $\approx 1000$ \\
        \hline
        \end{tabular}
    \label{tabble-parameter-verview}
\end{table}

\subsection{Deceleration and magneto-optical trap}\label{sec:deceltrap}
Molecules in a neon-cooled buffer-gas beam typically have a forward velocity of \SI{190}{m/s}~\cite{Mooij_2024}, so they need to be slowed down to be trapped by a MOT, whose capture velocity is below \SI{10}{m/s}~\cite{Augenbraun_2021a, Vilas_2022}. While cryogenic buffer-gas sources can create beams with a forward velocity of 30--\SI{50}{m/s}, this can only be done at the cost of molecule number~\cite{Lu_2011,White_2024}. Multiple ways to slow beams have been demonstrated, such as white-light and chirped-laser slowing~\cite{Vilas_2022, Anderegg_2023, Lasner_2024}. 

Here, we suggest a traveling-wave Stark decelerator, a device which leverages the Stark effect of low-field-seeking molecules to trap them in a co-moving frame which is slowed down gradually as the molecules pass through~\cite{van_de_Meerakker_2012}. In our own previous experiments, we have already demonstrated slowing SrF from \SI{190}{m/s} to standstill~\cite{Aggarwal_2021}. Our initial estimates suggested that up to $10^6$ BaF molecules per pulse can be slowed~\cite{Aggarwal_2018}, though with SrF, we have not been able to reach that goal. Still, many technical improvements can be made, such as transverse laser-cooling and focusing of molecules into the decelerator. For example, the ACME collaboration has demonstrated a factor 16 gain in ThO flux by using an electrostatic hexapole lens, and our own experiments with BaF in a similar setup suggest a factor 5 improvement~\cite{Wu_2022, Touwen_2024}. In addition, the acceptance velocity of the decelerator can be increased significantly by increasing its field strength and using molecules in the rotational state $N=2$. These methods should work even better for BaOH because its Stark shift is linear over a wide field range (see Fig.~\ref{fig-stark-curves}), making electrostatic focusing and deceleration easier~\cite{Touwen_2024}. We therefore expect that slowing $10^5$ BaOH molecules per pulse to the MOT capture velocity is possible.

As molecules leave the source, they are distributed over many internal states. Before the focusing and deceleration begin, optical pumping can be used to increase the yield of the desired state $\Tilde{X}^2\Sigma^+, (010), N=1, J=3/2^-$, which is low-field seeking at intermediate electric fields and therefore suitable for Stark deceleration.

We expect that the MOT cooling scheme will be similar to those described in \cite{Augenbraun_2023, Lasner_2022, Vilas_2022, Lasner_2024, Zeng_2024}, with $\Tilde{X}^2\Sigma^+ (000), N=1 \leftrightarrow \Tilde{A}^2\Pi_{1/2}, (000), J=1/2$, which is situated at \SI{870.81}{nm}, as the cooling transition and the repumpers predominantly addressing the $\Tilde{B}^2 \Sigma^+$ manifold to avoid reducing the scattering rate. A calculation and analysis of the Franck--Condon factors of the relevant transitions will be presented in a forthcoming publication. We are also considering a $\Tilde{X}(010) \leftrightarrow \Tilde{A}(010)$ transition for MOT cooling, which would have the advantage of reducing the number of optical pumping stages, since the molecules are needed in a (010) state before and after cooling in the MOT. To our knowledge, this has never been attempted before, but there is no fundamental reason it should not work.

\subsection{Optical transport}\label{sec:transport}
Because the MOT, which requires substantial $B$-field gradients, can't be easily situated inside the magnetically shielded volume, it is necessary to subsequently transport the molecules in a way which does not require $B$ fields. The method of launching molecules from the MOT via laser-frequency chirps~\cite{Fang_2009, Solmeyer_2013} is not easily adaptable to the case of BaOH with its complex level structure. We instead propose optical transport. 

There are three important challenges which need to be addressed: First, considering the thickness of the magnetic shielding ($\sim \SI{70}{cm}$), the transport distance must be rather long, which limits the achievable beam size. Second, BaOH is relatively heavy, such that large longitudinal trap frequencies are necessary to accelerate it. Third, the lowest reachable temperature for laser-cooled molecules is still high compared to the typical depth of an optical dipole trap. All of these problems can in principle be solved with more laser power, but this becomes impractical at some point. A better option is to use an optical lattice formed by two counter-propagating beams, one of which is a Bessel beam and therefore diverges more slowly over distance than an equivalent Gaussian beam. By modulating the phase of one beam in a sawtooth pattern, the lattice planes are moved along the beams, taking the molecules with them. This method has previously been demonstrated with cesium atoms~\cite{Klostermann_2022}, which have a comparable mass and ac polarizability to BaOH molecules, so a similar setup is likely to work in our case. Extrapolating from the parameters given there, 50\% of the loaded molecules can be transported and a total laser power of \SI{150}{W} is sufficient. This transport can be done horizontally, but it is preferable to set it up vertically (along the $Z$-direction). In this way, the steep longitudinal potential walls between the lattice planes not only accelerate the molecules for transport, but also support them against gravity. The transport can be completed in less than \SI{100}{ms}, keeping loss small and duty cycle high.

As explained in Section \ref{sec-spin-precession}, it may be necessary to load the molecules into a region of the science lattice which is larger than the extent of the transport trap along the $Y$-direction. This could be achieved by allowing the cloud to expand freely for a few milliseconds before ramping the science lattice on. Alternatively, one could gain more control over this process by piezo-mounting one of the lattice mirrors so that the laser frequency can be chirped while keeping the cavity in lock. The resulting longitudinal movement of the planes of the science lattice can then be used to distribute molecules during the loading process. However, we would like to avoid this because it introduces additional sources of trouble as discussed in the next section. After loading, the transport lattice must be turned off because its polarization can't be pure enough to avoid problematic vector light shifts.

\subsection{Science cavity}\label{sec:cavity}
The EDM measurement will take place in an optical lattice created by a laser beam inside a cavity. The dipole-trap potential inside the cavity takes the form of a number of circular planes, often called ``pancakes'', orthogonal to the cavity axis. This design solves two important problems simultaneously: first, it uses available laser power more efficiently, enabling a larger trap volume at a given laser power. This is crucial to allow large molecule numbers while keeping the density low. Second, by having the laser light pass through a polarization-filtering element inside the cavity, circularly polarized light can be removed much better. As previously demonstrated, this allows reaching a ratio of circular to linear polarization of $10^{-8}$~\cite{Zhu_2013a}. This is imperative for sufficient suppression of the vector light shift caused by the trap, since our calculations predict $\alpha^V(1064\,\text{nm})/(2hc\epsilon_0)=  6.6(4)~\text{Hz}/(\text{W}/\text{cm}^2)$, almost a quarter of the magnitude of the scalar component. 

The transverse size of the cavity mode can be increased by using cavity mirrors with a larger radius of curvature, however at some point the cavity mode becomes unstable and extremely sensitive to errors such as off-axis incoupling and imperfect surface parallelism of the spacer~\cite{Heinz_2021}. As demonstrated there, making a cavity with \SI{5}{cm} length and a mode diameter of \SI{1}{mm} is possible. Here, we suggest a 10-cm long cavity instead, to have sufficient space for transport and polarization filtering.

The use of a cavity introduces a number of challenges. The cavity must be sufficiently stable to avoid heating the molecules out of the trap due to the shaking motion of the pancakes. This is a resonant process which happens at the longitudinal trap frequency
\begin{equation}
\omega_y = \sqrt{\frac{2 \pi^2 U}{a^2 m}}.
\end{equation}
Here, $a = \SI{532}{nm}$ is the lattice spacing and $m = \SI{154.3}{amu}$ is the molecule mass. For the planned trap depth $U = h \times \SI{10}{MHz}$, this results in $\omega_y \approx 2 \pi \times \SI{220}{kHz}$. According to the estimation given in Ref.~\cite{Gehm_1998}, if we require the temperature-doubling time to be \SI{10}{s}, the spectral density of shaking at this frequency should be below $\SI{e{-13}}{\meter\per\sqrt{\hertz}}$.

There are two ways to achieve this kind of stability. The first is to use active feedback to control the mirror position and keep the cavity locked to be resonant with the laser. This makes the cavity design simple because its intrinsic stability matters less, but the mechanical feedback is limited by the weight of the mirrors. This makes it difficult to reach large lock bandwidths, resulting in larger noise outside the bandwidth. Despite this, an experiment with Cs atoms in a lattice with similar parameters as envisioned here has shown a lifetime of \SI{8}{s}~\cite{Zhang_2021}. Another potentially more serious problem is that the piezo element has to be mounted near the molecule cloud, such that any stray electromagnetic fields created by it can disturb the measurement. The second approach, which we choose here, is to build a very stable cavity and lock the laser to it. This offers extremely high stability, enough to reach more than \SI{100}{s} lifetime as demonstrated with Sr atoms~\cite{Heinz_2021, Park_2022}.

In the inset of Fig.~\ref{fig-setup}, we show the conceptual design of our spacer which offers a good compromise between the different requirements. The spacer is made of a single block of ultralow-expansion glass with apertures for light and molecules to pass through. It allows getting the electrodes close together and mounting them in an extremely accurate and stable way. It also contains an in-cavity Brewster plate to ensure polarization purity. One potential caveat of this design is that there are dielectric surfaces inside vacuum near the molecules. These can create uncontrolled electric-field offsets due to surface charges, which must be carefully characterized in the experiment.

We plan to use a cavity with a moderate finesse of approximately 1000, equivalent to a power-buildup factor of 300. Considering that we need a trap depth of at least $k_B\times\SI{500}{\micro\kelvin}$ in a lattice with a transverse area of $\approx \SI{1}{mm^2}$, this buildup factor brings the necessary laser power down to a reasonable \SI{5}{W}. It also limits high-frequency intensity noise in the cavity; considering that parametric heating occurs at twice the trap frequency, this may be helpful in limiting the heating rate caused by this process~\cite{Gehm_1998}. Intensity noise has little effect on the spin-precession coherence time because the intensity difference experienced by an average molecule as it travels through the trap is much larger than this noise. It was shown in Ref.~\cite{Edmunds_2013} that a high-power laser can be locked to a cavity with a similar finesse, and that trapping is not disturbed by thermal effects caused by the large circulating laser power. Using a higher finesse would reduce the requirement for laser power but also make the cavity mode narrower in frequency, which increases parametric heating caused by the conversion of frequency noise into amplitude noise~\cite{Heinz_2021, Gehm_1998}. 

\subsection{Creation of dc electric and magnetic fields}
\label{sec-implementation-fields}
Without any shielding, the background field in our lab is \SI{22}{\micro\tesla} with a noise level of \SI{500}{\pico\tesla} averaged over \SI{1}{s}. At the most important frequency scale of \SI{1}{Hz}, passive magnetic shielding made out of multiple layers of high-permeability material can reduce the field by a factor $10^6$ \cite{Liu_2021c, Ayres_2022}. Shielding of this type is commonly applied in eEDM experiments and is also currently used in our group~\cite{Boeschoten_2023, ACME_2017}. Extrapolating from this, it should be possible to reach a noise level of \SI{1}{fT} or less, which is significantly beyond what is needed for this experiment. At levels this low, the remanent magnetization of the shielding material will likely become the limiting factor.

To create a 0.1-nT background magnetic field with an rms noise of below \SI{4}{pT} in the frequency range around \SI{1}{Hz}, we propose to use a set of three Helmholtz coil pairs, one for each spatial direction. With this design, background field offsets in all directions can be cancelled out as well as three out of five independent components of the gradient. Additional coils with a more complicated shape can be added to also compensate the two remaining components of the gradient~\cite{Ang_2023}. There are enough degrees of freedom to align the resulting field to be orthogonal to the cavity axis, reducing the effect of vector light shifts as described in Section \ref{sec-optical-trapping}. The precision machining of the cavity spacer then automatically ensures that the $E$ and $B$ fields are mutually parallel.

Achieving a sufficiently stable electric field is more difficult because the acceptable noise is on the order of \SI{40}{\micro V/cm}, approximately $10^6$ times smaller than the field. Both time dependence and spatial inhomogeneity must be considered carefully. The electric field for polarizing the molecules along the $x$-axis will be created by two parallel plate electrodes made of ITO-coated glass. Great care must be taken to ensure parallelism, because field gradients over the \SI{130}{\micro\meter} extent of the molecule sample would compromise the experiment. At a distance of \SI{2.5}{cm} between the electrodes, an angle error of \SI{e{-4}}{rad} can be tolerated. This can be achieved by optically contacting the electrodes to the cavity spacer, which can be machined to micrometer-level precision. In addition, we need to ensure that the electrodes retain good enough surface flatness even after optical contacting. Keeping the $E$ field constant in time is also not an easy task. With one second averaging time, ultrastable voltage references based on ovenized Zener diodes can reach this level of stability~\cite{Beev_2022}. Very slow, carefully tuned feedback can transfer this stability to the electrodes. The limiting factors are likely to be thermal EMF and drift of electronic components, which can be minimized by careful choice of materials and components in combination with precise temperature control.

\subsection{Spin precession in an optical lattice}
\label{sec-spin-precession}
To perform the spin precession in a state with reduced magnetic moment but high state EDM sensitivity (see Section \ref{sec-field-stability}), the molecules are prepared in the state $\Tilde{X}^2\Sigma^+ (010), N=1, J=1/2^-, F=1$, with half of them in the lower Stark doublet where $m_{F}\ell = +1$, the other half in the upper doublet where $m_{F}\ell = -1$. Then, a coherent superposition between the $m_F=\pm1$ states should be created. More details on state preparation methods can be found in refs.~\cite{Leanhardt_2011, Caldwell_2023, Ho_2023}.

At a typical temperature of \SI{50}{\micro\kelvin}, a trap depth of $k_B \times \SI{500}{\micro\kelvin}$ and a lattice spacing of \SI{532}{nm}, the motion of molecules can be regarded as nearly classical, with an amplitude of \SI{55}{nm} in the longitudinal direction. This is much larger than the scattering length, so the collisional behavior can be approximated as three-dimensional with an effective volume of $\sim \SI{5e-9}{\cubic\centi\meter}$ per lattice plane. In other words, to hold the desired molecule number of 1300 (after transport) at an acceptable density of \SI{e{9}}{\per\cubic\centi\meter} (see Sec. \ref{sec-lifetime}), we should populate at least 250 planes, corresponding to a longitudinal cloud size of \SI{130}{\micro\meter}. 

Dipole-dipole interactions between neighbouring trapped molecules can also cause broadening of the spin-precession frequency, limiting the density. However, in this experiment there will always be an equal number of molecules with parallel and antiparallel alignment between molecular axis and external $E$ field, such that the shift cancels out in the mean-field approximation.

\subsection{Detection}\label{sec-detection}
A key advantage of working with a laser-coolable molecule is imaging. In the limit of few photons being detected per molecule, the statistical sensitivity of the experiment scales with the number of total detected photons because the probability to detect each molecule is small. We can instead operate in the regime where multiple photons are detected per molecule, such that almost every molecule contributes to the statistics. Then, the sensitivity is limited by the shot noise of the molecule number itself. Even with a small subset of the repumpers used in the MOT, each BaOH molecule can scatter 200 photons without significant optical pumping into dark states. This latter requirement is important because if a substantial fraction of molecules end up in a dark state during the measurement, the shot-noise limit can't be achieved~\cite{Lasner_2018}. Using an objective with a numerical aperture of 0.5 (i.e., covering 7\% of the solid angle) and an EMCCD camera or photomultiplier with a quantum efficiency of 25\%, we can detect 3 photons per molecule on average, giving a near-unit detection efficiency. If positioning an objective with such high numerical aperture inside the magnetically-shielded region turns out to be difficult, techniques like $\Lambda$-enhanced grey-molasses cooling~\cite{Cheuk_2018} could be used to scatter thousands of photons per molecule without losing them from the dipole trap. It has also been suggested to use entanglement to go below the shot-noise level~\cite{Aoki_2021, Zhang_2023a}, but this is difficult to do with moderate particle numbers even in the much simpler case of atoms and is beyond the horizon for this proposal. 

\section{Conclusion}\label{sec-conclusion}
We have investigated the experimental requirements and constraints for a future eEDM measurement with neutral and open-shell polyatomic molecules trapped in an optical lattice. This is an attractive idea because it enables long coherence times, large molecule numbers and small trapping volumes, allowing high statistical sensitivity and reduced influence of spatial inhomogeneity of external fields. However, many obstacles remain on the path to a state-of-the-art eEDM measurement with polyatomics. Better molecule sources, decelerators, MOTs and optical-trap loading schemes need to be developed to reach the required molecule numbers. Careful engineering has to go into the design of the science cavity, magnetic shielding, magnetic-field coils and many other parts of the experiment. The most critical aspects identified here are the homogeneity and stability of the applied fields as well as the polarization purity of the trapping laser. If these obstacles can be overcome, polyatomic molecules may become the basis of the next generation of eEDM measurements.

\section{Outlook}\label{sec-outlook}
The most obvious improvement to the proposal outlined here is increasing the molecule number. In principle, up to $10^5$ molecules could be trapped in an optical lattice if the available volume were used efficiently, increasing statistical sensitivity by an order of magnitude. The main limiting factors are currently molecule sources and MOTs, which can't produce large and cold enough samples. However, it is reasonable to assume that improvements can be made: for example, the first time diatomic molecules were loaded from a MOT into an optical dipole trap, the number was on the order of 100~\cite{Anderegg_2018}, but within a few years, it was increased to 26000 at a comparable temperature~\cite{Yu_2024}. Another idea is to leverage the much longer lifetimes in optical traps compared to MOTs, loading multiple molecule samples from the MOT into different positions along the science lattice before spin precession. This requires smart hand-over procedures between dipole traps, but is possible in principle and could yield a five- to ten-fold number improvement. There are also ways to increase the lifetime to allow longer $\tau$ and smaller loss. For example, switching to a symmetric-top molecule like BaOCH$_3$ allows at least one order of magnitude increase in lifetime against spontaneous decay as explained in the supplementary information of Ref.~\cite{Kozyryev_2017a}. In combination with a cryogenically-cooled vacuum chamber to reduce blackbody-radiation loss and a further-detuned dipole trap, $\tau \approx \SI{10}{s}$ may be reachable, allowing another order of magnitude gain in statistical sensitivity.

So, there is hope for improving the shot-noise-limited performance, but reaching this limit may pose a bigger problem due to field noise. In this proposal, we assume that the $E$ field can be controlled to a relative precision of $10^{-6}$. While further improvements are certainly possible, the limits of precision machining, control of temperature-dependent effects, and shielding against electromagnetic interference are not far away. Therefore, to get the full potential out of an experiment with larger molecule numbers, we need to find molecules which are less sensitive to field noise.

\section*{Acknowledgments}
We thank Yuly Chamorro, Scott Eustice, Thomas Meij\-knecht, Odile Smits, Michael Tarbutt, Anno Touwen, and Julian Wienand for helpful discussions. The work of LFP, NB and SH is supported by project number VI.C.212.016 financed by the Dutch Research Council (NWO). The work of SH, AB, TF and IAA is supported by the ENW-M2 NWO grant OCENW.M.21.098. The work of SH, AB and EP is supported by ENW-XL  NWO grant OCENW.XL21.XL21.074. IAA acknowledges partial support from FONCYT through grants PICT-2021-I-A-0933 and PICT-2020-SerieA-0052, and from CONICET through grant PIBAA-2022-0125CO.

\bibliography{bibliography}

\end{document}